# SARG control system

R. COSENTINO[1], G. BONANNO[1], P. BRUNO[1], A. CALi[1], S. SCUDERI[1], M. C. TIMPANARO[1]

[1]*Osservatorio Astrofisico di Catania, Catania, Italy*

ABSTRACT. The control system and the entire architecture of the High Resolution Spectrograph (SARG) for the Italian National Telescope "Galileo" (TNG) are here described. The concept of SARG instrument controls is similar to that of the other TNG instruments, in particular the CCD detector driving and the image acquisition use the same TNG standard boards and the same selected bus: the VME. The link between the SARG VME and the other telescope components is based on the same GATE software that guarantees the compatibility with the entire distributed TNG software. The control of the moving parts as well as the other parts of instrument that is the lamp controller and the temperature sensors, is based on a commercial controller connected to the system through a serial link. Furthermore a specialized software running on a PC has been realized to test the rotating tables independently of the VME system. Test of accuracy and repeatability of the positioning were done and some results are presented.

## 1. SARG Detectors and Instrument Control Architecture

The SARG instrument controls use the same approach adopted for the other instruments installed at the TNG, as for example the Optical Imager Galileo (OIG). The CCD detector driving and the image acquisition use the same TNG standard boards and the same selected bus: the VME. The link between the SARG VME and the other telescope components is based on the same GATE software that guarantees the compatibility with the entire distributed TNG software. In figure 1 is shown the block diagram of the entire system.

As can be derived from this figure two sets of VME boards (Manufactured by ATENIX) provide the link to the CCD controller used to drive the scientific detector and to the controller that drives the CCD of the slit viewer. The two ATENIX boards use also the VSB bus to manage efficiently the available memory. A serial interface standard RS-232 is used to control the entire SARG mechanical system. The connection between the serial port of the CPU and the serial port of the various controllers of the instrument is made by using a serial multiplexer (BAYTECH) that is able to switch the serial communication to four different serial devices. These serial devices are the controllers:

1. Motor Controller 1
2. Motor Controller 2
3. Lamp Controller
4. Temperature Monitoring

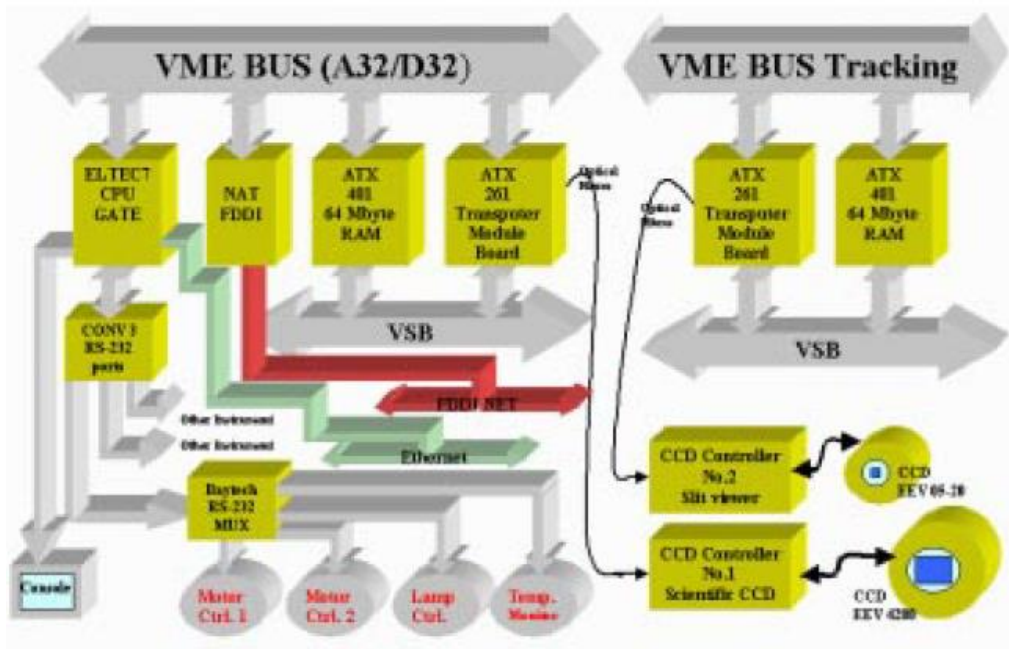

Fig. 1. SARG detectors and instrument conrol architecture.

## 2. Instrument Control

The instrument control architecture is based essentially on the RS-232 standard communication protocol. We select the various controllers equipped with the RS-232 standard interface. Through the serial link and the associated GATE software task we are able to drive the various controllers. We use three motor controllers, a lamp controller and a temperature monitoring. Figure 2 shows the various movable parts, the lamp controller and the temperature monitoring that constitute the entire control system.

A motor controller equipped with four motor drivers is able to control:

1a)  the CALIBRATION MIRROR SLIDE (**CLS**)

2a)  the PRESLIT SLIDE (**PS**)

3a)  the LAMP SELECTION TABLE (**LST**)

4a)  the SLIT VIEWER SLIDE (**SVS**)

A motor controller equipped with other four motor drivers is able to control:

lb)  the FILTER WHEEL (**FW**)

2b)  the SLIT WHEEL (**SW**)

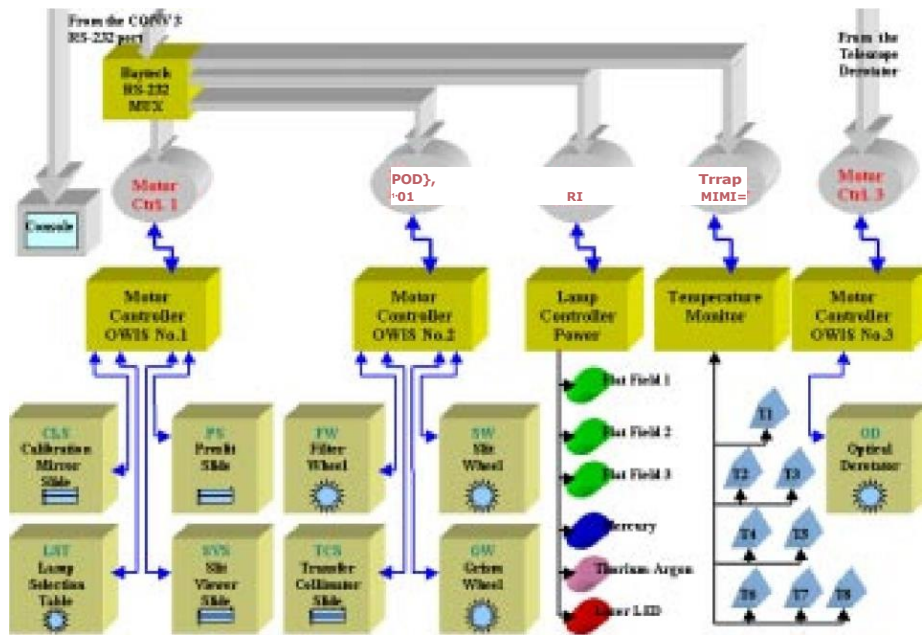

Fig. 2. Instrument control architecture.

3b) the GRISM WHEEL (**GW**)

4b) the TRANSFER COLLIMATOR SLIDE (**TCS**)

A motor controller equipped with just one motor driver is dedicated to control the OPTICAL DEROTATOR (**OD**).

The lamp controller is able to power up the three Flat Field lamps, the two calibration lamps and a laser LED, while the temperature monitoring is able to measure the temperature in 8 different places of the instrument.

To have an idea of how the various controlled parts are connected with the CENTRAL CONTROL UNIT constituted physically by a RACK that mount the various controllers in Figure 3 are showed the connections related to the control of the instrument and those related to the detectors are showed in Figure 4.

## 3. Motor Controller and Movements

The selected motor controller to drive all the movement of the instrument is the DC-500 manufactured by OWIS. The controller is able to drive 6 independent axes. The commands can be sent by using either the keypad , either the serial interface standard

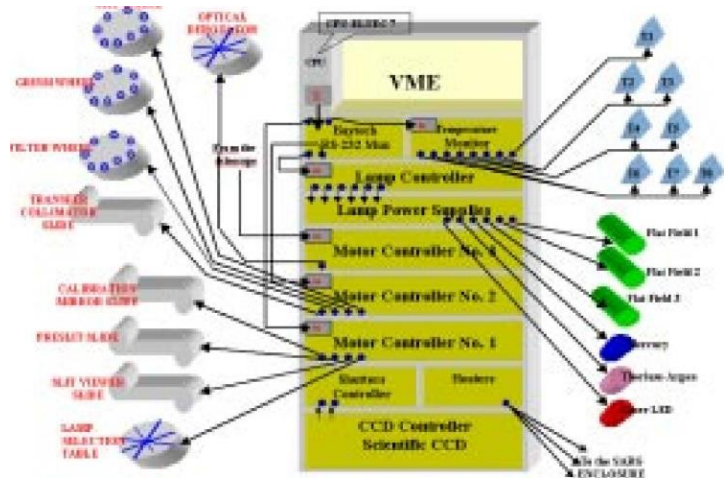

Fig. 3. SARG Motor lamps and temperature monitoring electronics controller rack.

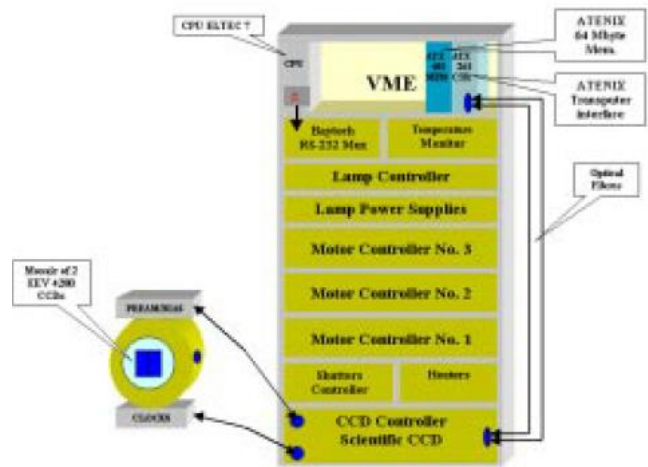

Fig. 4. SARG CCD electronics controller rack.

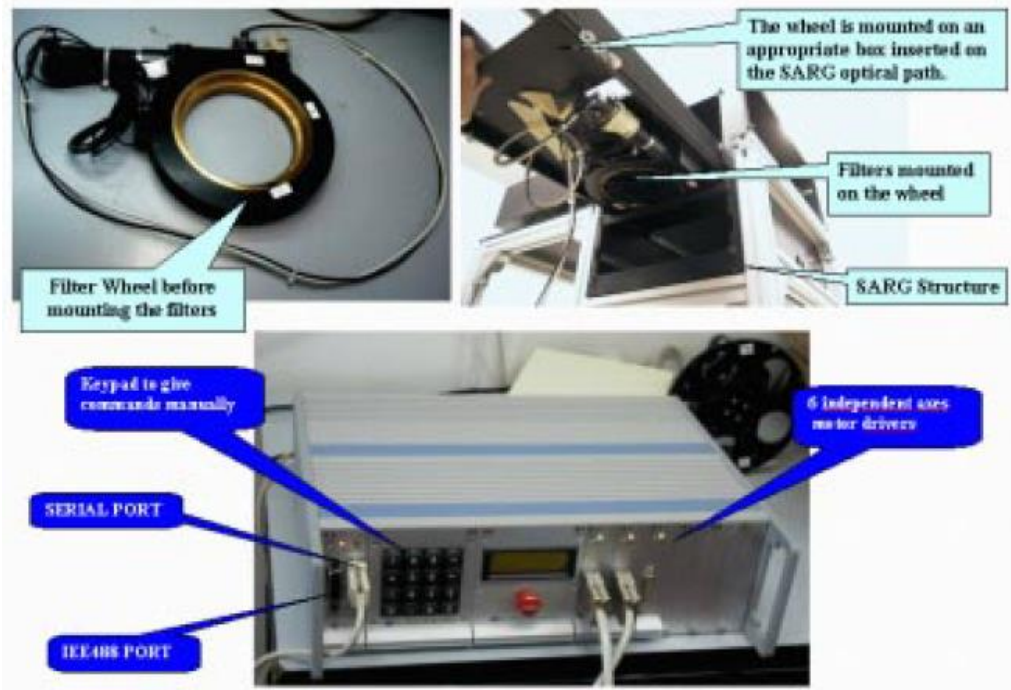

Fig. 5. SARG OWIS Motor controller and the filter wheel before the mounting on SARG (upper left panel) and during the mounting inside the SARG (upper right panel).

RS-232 or the IEE488 parallel port. This is useful to test immediately the rotating or linear movements by using the function allowed by the keypad, as for example the zero search, or the go and stop of the movement. The serial interface, allows to implement an easy comunication link with a PC, and thus a total control through dedicated software procedures. This last is very useful during the debug and maintenance of the instrument. Infact we can switch the serial link from the VME crate to the PC and avoid the 'TNG complex net. Figure 5 shows an example of wheel (the filter wheel).

The wheel before mounting the filters is shown on the left side of the panel while the wheel completed with the filter holders is shown on the right side of the panel. The wheel is mounted in an appropriate box and inserted along the SARG optical path. To note the ability to remove each single box for mechanical maintenance.

## 4. Software

The RS-232 serial standard interface allows us to develop a VISUAL BASIC procedure running on a PC under Windows 95 Operating System. This approach is simpler than the complete system that use the VME crate and the GATE software. The procedure is able to perform all the required measurements to check the mechanical movements reliability. Figure 6 shows how the procedure after the initialization presents some windows that allow various usefull commands to drive the motor controller.

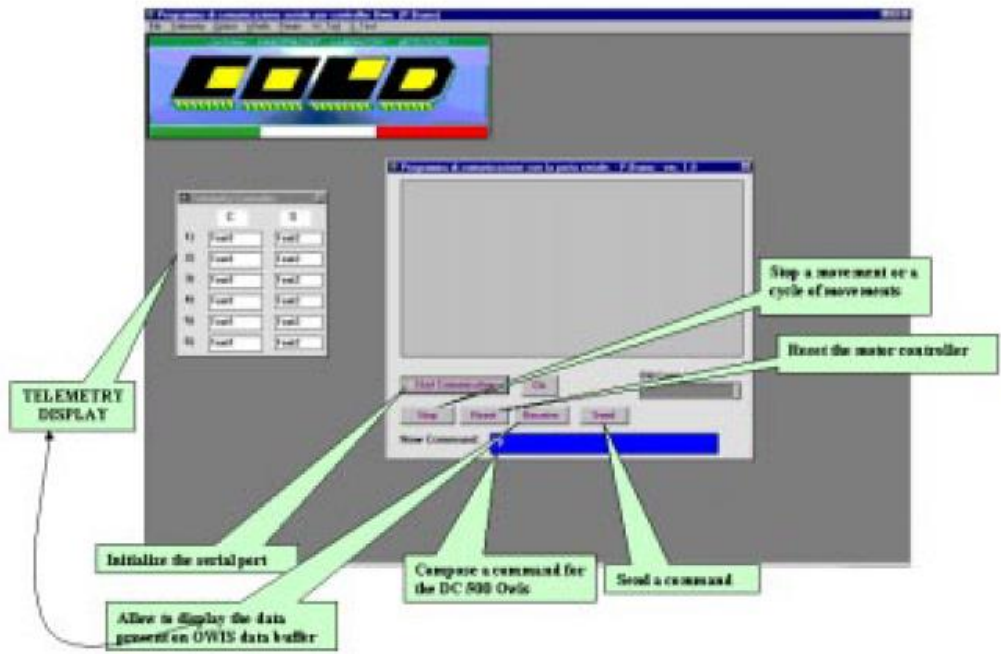

Fig. 6. Visual Basic procedure (for WIN 95/98) to test the wheels.

Through this procedure it is possible to initialize the serial port, compose and send the commands to the DC500 motor controller and display the telemetry data. Furthermore we are able to change some parameters of each controlled axis and execute the tests. In figure 7 is shown the user interface with all the procedures to drive the DC-500 motor controller and to allow the tests of the movements and the measurements of the positions.

## 5. Tests and Results

In order to check the accuracy of the positon and their repeteabilty over a certain number of movements we have done some measurements on all the rotators and the linear actuators.

In figure 8 the deviation from the zero position is plotted versus the number of measurements for a rotating table that moves a 20 N load.

Each single measure is obtained as follows:

a) move the table to the zero reference and clear the increment counter;

b) move the table to a quarter of turn and move back to zero position;

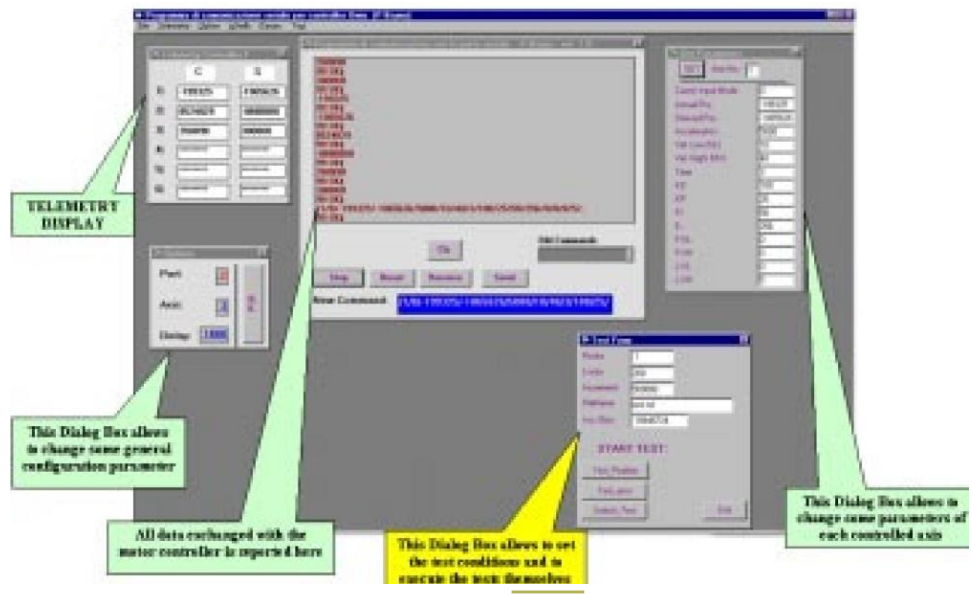

Fig. 7. All available procedures to drive the OWIS DC 500 motor controller and test the wheels.

c) save in a file the displacement (deviation) respect to the zero position.

As can be seen from the plot the deviation from the zero position is, on average, 5.8 increments corresponding to 1". The standard deviation of the data comes out to be 23.38 increments or 3.5".

References

Gratton, R.G., Claudi, R.U., Rebeschini, M., Bonanno, G., Bruno, P., Cali, A., Scuderi, S., Cosentino, R., Desidera, S. : 1998, in *The High Resolution Spectrograph of TNG: a Status Report, Techical* report n 72.

Gratton, R.G., Claudi, R.U., Rebeschini, M., Martorana, G., Farisato,G.,Bonanno, G., Bruno, P., Cali, A., Scuderi, S., Cosentino, R., Desidera, S. : 1999, in *The High Resolution Spectrograph of TNG,* The proceedings of this conference.

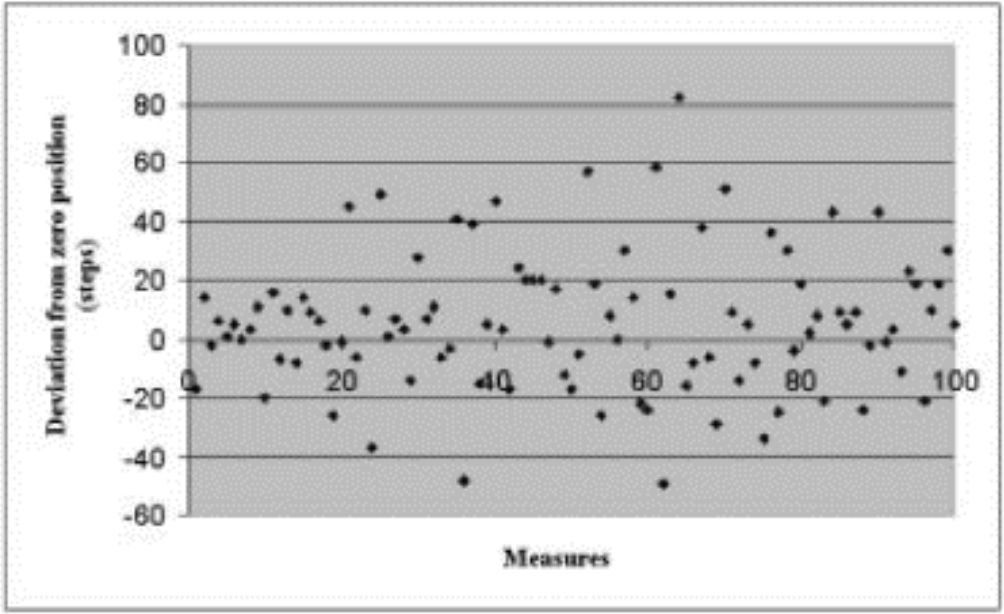

Fig. 8. Deviation from the zero position plotted versus the number of measurements for a rotating table that moves a 20 N load. The table is in vertical position.